\documentclass[10pt]{article}

\usepackage{amssymb}

\newcommand{\beq}{\begin{equation}}
\newcommand{\eeq}{\end{equation}}
\newcommand{\bea}{\begin{eqnarray}}
\newcommand{\eea}{\end{eqnarray}}

\begin{document}
\pagestyle{empty}

\hfill hep-th/0603188

\begin{center}

\vspace*{50mm}
{\LARGE Wilson-'t~Hooft operators and the theta angle}

\vspace*{20mm}
{\Large M{\aa}ns Henningson}

\vspace*{10mm}
Department of Fundamental Physics\\
Chalmers University of Technology\\
S-412 96 G\"oteborg, Sweden

\end{center}

\vspace*{20mm} \noindent
{\bf Abstract:}\\
We consider $(3 + 1)$-dimensional $SU (N) / \mathbb Z_N$ Yang-Mills theory on a space-time with a compact spatial direction, and prove the following result: Under a continuous increase of the theta angle $\theta \rightarrow \theta + 2 \pi$, a 't~Hooft operator $T (\gamma)$ associated with a closed spatial curve $\gamma$  that winds around the compact direction undergoes a monodromy $T (\gamma) \rightarrow T^\prime (\gamma)$. The new 't~Hooft operator $T^\prime (\gamma)$ transforms under large gauge transformations in the same way as the product $T (\gamma) W (\gamma)$, where $W (\gamma)$ is the Wilson operator associated with the curve $\gamma$ and the fundamental representation of $SU (N)$. 

\newpage \pagestyle{plain}
\section{Introduction}
Wilson operators and 't Hooft operators constitute important observables in non-abelian Yang-Mills theory in $d = 3 + 1$ dimensions. In this paper, we will consider the gauge group 
\beq
G \simeq SU (N) / C , 
\eeq
where $C \simeq \mathbb Z_N$ denotes the center of $SU (N)$. The basic Wilson operator $W (\gamma)$ associated with a closed spatial curve $\gamma$ is then defined as
\beq
W (\gamma) = \frac{1}{N} {\rm Tr} \left( P \exp \int_\gamma A \right) , \label{W}
\eeq
where $A$ is the connection one-form, $P$ denotes path ordering along $\gamma$, and ${\rm Tr}$ is the trace in the fundamental representation of $SU (N)$. The operator $W (\gamma)$ is invariant under gauge transformations that can be continuosly deformed to the identity transformation. Under a general gauge transformation, $W (\gamma)$ is multiplied by an $N$-th root of unity determined by the class in $\pi_1 (G) \simeq \mathbb Z_N$ of (the restriction to $\gamma$ of) the gauge transformation. 

The definition of the corresponding 't~Hooft operator $T (\gamma)$ is less explicit \cite{'tHooft}:  On the complement of $\gamma$ in space, $T (\gamma)$ is given by a $G$ valued gauge transformation, whose restriction to another closed curve that links $\gamma$ once represents the image of the generator $1$ of $\mathbb Z_N$ under the isomorphism $\mathbb Z_N \simeq \pi_1 (G)$. Such a transformation has a well-defined action on all the fields of the theory, but is obviously singular at the locus of $\gamma$. By deforming the transformation over a tubular neighbourhood of $\gamma$, we regularize it to a smooth transformation defined over all of space-time.  The precise form of the regularization is of no consequence for the arguments of the present paper; the important point is that the resulting field configuration is smooth everywhere. The regularized transformation is however not a gauge transformation, so the 't Hooft operator $T (\gamma)$ thus defined has a non-trivial action also on gauge invariant states. 

This definition of the 't~Hooft operator in terms of a singular gauge transformation is ambigious in the sense that it allows for the multiplication of $T (\gamma)$ by a phase-factor, which may be an arbitrary gauge-invariant functional of the fields of the theory. It may even be impossible to give a globally valid prescription for fixing this ambiguity. Indeed, it has been stated in several papers (see for example \cite{Bianchi-Green-Kovacs}\cite{Cachazo-Seiberg-Witten}\cite{Kapustin}), that under a smooth increase $\theta \rightarrow \theta + 2 \pi$ of the theta angle, $T (\gamma)$ undergoes a monodromy
\beq
T (\gamma) \rightarrow T^\prime (\gamma) , \label{monodromy} 
\eeq
where the new 't~Hooft operator $T^\prime (\gamma)$ behaves as the product $T (\gamma) W (\gamma)$; it could be called a Wilson-'t Hooft operator. (On the other hand, the explicit expression (\ref{W}) shows that the corresponding monodromy of the Wilson operator $W (\gamma)$ is trivial; $W (\gamma) \rightarrow W (\gamma)$.) This monodromy of operators associated with closed spatial curves is analogous to  the Witten effect \cite{Witten78}, which amounts to an increase of the electric charge of a magnetically charged dyonic particle state as $\theta \rightarrow \theta + 2 \pi$ continuously. 

However, we are not aware of any published proof of the monodromy transformation (\ref{monodromy}). The aim of the present paper is to provide such a proof, based on a topological obstruction that prevents a global definition of $T (\gamma)$. The obstruction will only be present if the curve $\gamma$ represents a non-trivial homotopy class. We will therefore consider the theory on a spatial three-manifold $X$ of the form
\beq
X \simeq S^1 \times \mathbb R^2 , 
\eeq
and let $\gamma$ wind once around the $S^1$ factor of $X$. However, even if there is no topological obstruction against a global definition of the 't~Hooft operator if we take the spatial manifold as $\mathbb R^3$, it is certainly natural to assume a similar monodromy transformation also in this case.

In the next section, we will review the interpretation of the 't~Hooft operator $T( \gamma)$ in terms of the topology of principal $G$ bundles over space.  In section three, we will consider the topology of the group of gauge transformations. A gauge transformation may be winded in two different ways: along the curve $\gamma$, or over three-space as a whole. As discussed above, gauge transformations that are winded along $\gamma$ have a non-trivial action on the Wilson operator. The transformation properties under gauge transformations that are winded over three-space as a whole are described by the theta angle. In section four, we will show how the interplay of these effects leads to the monodromy property (\ref{monodromy}).

\section{'t~Hooft operators}
We begin by reviewing the classification of principal $G \simeq SU (N) / C$ bundles over a low-dimensional compact connected space $B$. This follows from the first few homotopy groups of $G$:
\beq
\pi_i (G ) \simeq \left\{ \begin{array}{ll} 0,  & i = 0 \cr \mathbb Z_N, & i = 1 \cr 0, & i = 2 \cr \mathbb Z, & i = 3 . \end{array} \right.
\eeq
Thus, for a one-dimensional base space $B$, all $G$ bundles are trivial. For a two- or three-dimensional $B$, they are classified by a characteristic class 
\beq
w_2 \in H^2 (B, \mathbb Z_N) ,
\eeq
known as the second Stiefel-Whitney class in mathematics or the discrete magnetic flux in physics. For a four-dimensional $B$ there is an additional characteristic class
\beq
c_2 \in H^4 (B, \mathbb Q) ,
\eeq
known as the second Chern class or the instanton number. It is related to the second Stiefel-Whitney class $w_2$ as
\beq
c_2 = \frac{1}{2} \left(\frac{1}{N} - 1 \right) \bar{w}_2 \cup \bar{w}_2 {\rm \; mod \;} H^4 (B, \mathbb Z) , \label{nonintegrality}
\eeq
where $\bar{w}_2 \in H^2 (B, \mathbb Z)$ denotes an arbitrary lifting of $w_2$ to an integral class. (An instructive proof of this relation can be found in e.g. \cite{Vafa-Witten}.) In higher dimensions, there are further invariants, but they will not be needed in the present  paper.

Consider now a state of finite energy in Yang-Mills theory with gauge group $G$ on the spatial manifold $X \simeq S^1 \times \mathbb R^2$. As we go to infinity in the $\mathbb R^2$ factor, all physical data must approach their vacuum values. We may therefore add the points at infinity, thereby replacing $X$ by the compact space 
\beq
X^\prime \simeq S^1 \times S^2 . 
\eeq
However, while a $G$ bundle $P$ over $X$ is necessarily trivial (since $H^2 (X, \mathbb Z_N) \simeq 0$), this is not so for a $G$ bundle $P^\prime$ over $X^\prime$; according to the previous paragraph, such bundles are classified by a characteristic class $w_2^\prime \in H^2 (X^\prime, \mathbb Z_N) \simeq \mathbb Z_N$. 

It is now easy to understand the action of an 't~Hooft operator $T (\gamma)$ associated with a closed curve $\gamma$ that winds once around the $S^1$ factor of $X$ or $X^\prime$: When acting on a state $\left| \psi \right>$ with a definite value $w_2^\prime$ of the second Stiefel-Whitney class, it produces another state $| \tilde{\psi} > = T (\gamma) \left| \psi \right>$ for which the second Stiefel-Whitney class takes the value $\tilde{w}_2^\prime$ given by
\beq
\tilde{w}_2^\prime = w_2^\prime + 1.
\eeq
(In this formula, we identify a class in  $H^2 (X^\prime, \mathbb Z_N)$ with its image under the  isomorphism $H^2 (X^\prime, \mathbb Z_N) \simeq \mathbb Z_N$.)

\section{Large gauge transformations}
This section is largely inspired by \cite{Witten02}. 

Let $P^\prime$ be a $G$ bundle over $X^\prime \simeq S^1 \times S^2$, characterized by its value $w_2^\prime \in H^2 (X^\prime, \mathbb Z_N)$ of the second Stiefel-Whitney class, as described in the previous section. We let ${\cal G}$ denote the group of gauge transformations, i.e. the group of bundle automorphisms of $P^\prime$. It is not connected; the component of ${\cal G}$ containing the identity transformation is a normal subgroup, which we denote as ${\cal G}_0$. Physical states must be invariant under ${\cal G}_0$, but they need not be invariant under all of ${\cal G}$. Their transformation properties may be given  by an arbitrary character of the quotient group of homotopy classes of gauge transformations
\beq
\Lambda \simeq {\cal G} / {\cal G}_0 .
\eeq

We will need to understand the structure of the discrete abelian group $\Lambda$. Let $\Lambda_\gamma \simeq \pi_1 (G) \simeq \mathbb Z_N$ be the group of homotopy classes of gauge transformations for the trivial bundle over the spatial curve $\gamma$ that is obtained by restricting the bundle $P^\prime$ to $\gamma$. Let $\Lambda_0 \simeq \pi_3 (G) \simeq \mathbb Z$ be the subgroup of $\Lambda$ consisting of homotopy classes of gauge transformations of $P^\prime$ that are trivial when restricted to $\gamma$. We thus have a short exact sequence
\beq
0 \rightarrow \Lambda_0 \stackrel{i}{\rightarrow} \Lambda \stackrel{r}{\rightarrow} \Lambda_\gamma \rightarrow 0 ,
\eeq
where $i$ and $r$ are the obvious inclusion and restriction maps respectively. In other words, the group $\Lambda$ is an extension of $\Lambda_\gamma \simeq \mathbb Z_N$ by $\Lambda_0 \simeq \mathbb Z$. To describe this extension precisely, we choose a $\lambda \in  \Lambda$ such that $r (\lambda)$ equals the image of the generator $1$ of  $\mathbb Z_N$ under the isomorphism $\mathbb Z_N \simeq \Lambda_\gamma$. Since $\lambda^N \in {\rm \; ker \;} r$ and the sequence is exact,  $\lambda^N \in {\rm \; Im \;} i$, so 
\beq
\lambda^N = \Omega^k , 
\eeq
where $\Omega$ is the generator of $\Lambda_0 \simeq \mathbb Z$ and $k$ is some integer, which depends on the choice of $\lambda$. (Here we have switched to a multiplicative rather than additive notation for the group operations.) 

To compute the integer $k$, we consider the four-dimensional space
\beq
Y \simeq S^1 \times X^\prime \simeq S^1 \times S^1 \times S^2 \label{Y} .
\eeq
(As will become clear, this auxiliary space should not be thought of as a space-time.) We construct two $G$ bundles $P^\lambda$ and $P^\Omega$ over $Y$ by first extending the given bundle $P$ over the cylinder $I \times X^\prime$, where $I$ is an interval, and then gluing the ends together with gluing data $\lambda$ or $\Omega$ respectively. We then have that 
\beq
N c_2^\lambda = k c_2^\Omega ,
\eeq
where $c_2^\lambda$ and $c_2^\Omega$ are the second Chern classes of the bundles $P^\lambda$ and $P^\Omega$ respectively. The class $c_2^\Omega \in H^4 (Y, \mathbb Q)$ is given by the image of the element $1$ of $\mathbb Q$ under the isomorphism $\mathbb Q \simeq H^4 (Y, \mathbb Q)$. According to (\ref{nonintegrality}), the class $c_2^\lambda \in H^4 (Y, \mathbb Q)$ is determined modulo $H^4 (Y, \mathbb Z)$ by the second Stiefel-Whitney class $w_2^\lambda \in H^2 (Y, \mathbb Z_N)$ of $P^\lambda$. The latter class is determined by its restrictions to the factors $S^1 \times S^1$ and $S^2$ on the right hand side of (\ref{Y}). The restriction of $w_2^\lambda$ to $S^1 \times S^1$ is in fact given by $r (\lambda) \in H^2 (S^1 \times S^1, \mathbb Z_N) \simeq \Lambda_\gamma$. The restriction of $w_2^\lambda$ to $S^2$ equals the second Stiefel-Whitney class $w_2^\prime \in H^2 (S^2, \mathbb Z_N) \simeq H^2 (X^\prime, \mathbb Z_N)$ of $P^\prime$. Thus
\beq
w_2^\lambda = p_1^*(r (\lambda)) + p_2^* (w_2^\prime) ,
\eeq
where $p_1$ and $p_2$ are the projections from $Y$ to $S^1 \times S^1$ and $S^2$ respectively. A small calculation now gives
\beq
c_2^\lambda = \frac{1}{N} p_1^* (r(\lambda)) \cup p_2^* (w_2^\prime)  {\rm \; mod \;} H^4 (Y, \mathbb Z) .
\eeq
Putting everything together, we find that $k = w_2^\prime {\rm \; mod \;} N$, where $w_2^\prime$ denotes the image of the second Stiefel-Whitney class of $P^\prime$ under the isomorphism $H^2 (X^\prime, \mathbb Z_N) \simeq \mathbb Z_N$. (Since $\lambda$ is only defined modulo $\Omega$, we can only determine $k$ modulo $N$.)

In summary, we have found that the group $\Lambda$ is generated by the elements $\lambda$ and $\Omega$, subject to the relation
\beq
\lambda^N = \Omega^{w_2^\prime {\rm \; mod \;} N} . \label{relation}
\eeq

\section{The monodromy}
A physical state $\left| \psi \right>$ is characterized by a certain value $w_2^\prime$ of the second Stiefel-Whitney class, as described in section two. Its transformation properties under the discrete abelian group $\Lambda$ described in the previous section can be specified by the eigenvalues $e^{i \theta}$ and $e^{i \phi}$ of the generators $\Omega$ and $\lambda$ respectively:
\bea
\Omega \left| \psi \right> & = & e^{i \theta} \left| \psi \right> \cr
\lambda \left| \psi \right> & = & e^{i \phi} \left| \psi \right> .
\eea
As the notation suggests, $\theta$ is indeed the theta angle parameter of Yang-Mills theory. The relation (\ref{relation}) implies that
\beq
e^{i \phi N} = e^{i (w_2^\prime + n N) \theta}
\eeq
for some integer $n$. If we follow a particular solution to this equation under a continuous increase $\theta \rightarrow \theta + 2 \pi$, the eigenvalue $e^{i \phi}$ undergoes the monodromy
\beq
e^{i \phi} \rightarrow e^{i \phi} e^{2 \pi i w_2^\prime / N} .
\eeq

Acting with an 't~Hooft operator $T (\gamma)$ on $\left| \psi \right>$ produces another state $| \tilde{\psi} > =T (\gamma) \left| \psi \right>$ with the value $\tilde{w}_2^\prime = w_2^\prime + 1$ of the second Stiefel-Whitney class. Repeating the above argument, we find that the corresponding eigenvalue $e^{i \tilde{\phi}}$ of the generator $\lambda$ undergoes the monodromy
\beq
e^{i \tilde{\phi}} \rightarrow e^{i \tilde{\phi}} e^{2 \pi i (w_2^\prime + 1) / N} .
\eeq

The different monodromy properties of the two states mean that the 't~Hooft operator must undergo a monodromy
\beq
T (\gamma) \rightarrow T^\prime (\gamma) .
\eeq
The quotient $\hat{W} (\gamma) = T^\prime (\gamma) T^{-1} (\gamma)$ can be characterized by its transformation property under $\lambda$:
\beq
\lambda  \hat{W} (\gamma) \lambda^{-1} = e^{2 \pi i / N} \hat{W} (\gamma) .
\eeq
But this agrees with the transformation property of the Wilson operator $W (\gamma)$ in the fundamental representation of $SU (N)$ as defined in (\ref{W}). So although the present arguments do not give an exact description of $T^\prime (\gamma)$ (which would depend on the precise prescription for regularizing the 't~Hooft operators in the vicinity of $\gamma$), we can conclude that $T^\prime (\gamma)$ indeed transforms in the same way as the product $T (\gamma) W (\gamma)$ under gauge transformations.

\vspace*{5mm}
I am supported by a Research Fellowship from the Royal Swedish Academy of Sciences.

\end{document}